# Freeform nanostructuring of hexagonal boron nitride

Nolan Lassaline[1], Deepankur Thureja[1,2], Thibault Chervy[2,3], Daniel Petter[1], Puneet A. Murthy[2], Armin W. Knoll[4], and David J. Norris[1]

[1]Optical Materials Engineering Lab, Dept. of Mechanical and Process Engineering, ETH Zurich, 8092 Zurich, Switzerland.
[2]Quantum Photonics Group, Dept. of Physics, ETH Zurich, 8092 Zurich, Switzerland.
[3]Physics and Informatics Laboratories, NTT Research Inc., 940 Stewart Drive, Sunnyvale, California 94085, USA.
[4]IBM Research - Zurich, 8803 Rueschlikon, Switzerland.

**Hexagonal boron nitride (hBN)—long-known as a thermally stable ceramic—is now available as atomically smooth, single-crystalline flakes[1], revolutionizing its use in optoelectronics[2-5]. For nanophotonics, these flakes offer strong nonlinearities[6], hyperbolic dispersion[4,7,8], and single-photon emission[9], providing unique properties for optical and quantum-optical applications. For nanoelectronics, their pristine surfaces, chemical stability, and wide bandgap have made them the key substrate[2,10,11], encapsulant[12,13], and gate dielectric[14] for two-dimensional electronic devices[15]. However, while exploring these advantages, researchers have been restricted to flat flakes or those patterned with basic slits[8] and holes[16], severely limiting advanced architectures. If freely varying flake profiles were possible, the hBN structure would present a powerful design parameter to further manipulate the flow of photons[7], electrons[14,15], and excitons[17,18] in next-generation devices. Here, we demonstrate freeform nanostructuring of hBN by combining thermal scanning-probe lithography[19,20] and reactive-ion etching to shape flakes with surprising fidelity. We leverage sub-nanometer height control and high spatial resolution to produce previously unattainable flake structures for a broad range of optoelectronic applications. For photonics, we fabricate microelements such as phase plates, grating couplers, and lenses. We show the straightforward transfer and integration of such elements by placing a spherical hBN microlens between two planar mirrors to obtain a stable, high-quality optical microcavity[21]. We then decrease the patterning length scale to introduce Fourier surfaces[22] for electrons, creating sophisticated, high-resolution landscapes in hBN, offering new possibilities for strain[17,18] and band-structure[14,16,23,24] engineering. These capabilities can advance the discovery and exploitation of emerging phenomena in hyperbolic metamaterials[25], polaritonics[26,27], twistronics[28], quantum materials[29], and deep-ultraviolet optoelectronic devices[1,30].**

Hexagonal boron nitride (hBN) is a layered crystal, like graphite and molybdenum disulfide, that can be exfoliated from thin flakes down to monolayer sheets while maintaining pristine, atomically smooth surfaces[2,3]. At nanoscale thicknesses, such crystals, known as two-dimensional (2D) materials, exhibit unique electronic and optical properties. Moreover, different 2D materials can be selected, stacked, and stuck together via van der Waals (vdW) forces[3,15], providing a flexible route to 'designer' materials for the discovery of unexpected phenomena and the creation of optoelectronic devices with unprecedented performance. These 'vdW heterostructures' continue to advance rapidly due to stacking innovations[31] (for example, twisted bilayers[11]) and a growing library of 2D materials. Yet, while the active layers and electrode materials in these devices can be varied, hBN has remained an irreplaceable component. First identified as the ideal substrate for graphene electronics[2], hBN is now ubiquitous because of its special ability to surround, protect, and isolate 2D materials. In addition, hBN is increasingly investigated for nanophotonic applications[5], offering infrared hyperlensing[4,7,8], room-temperature single-photon emitters[9], and bright deep-ultraviolet sources[1,30].

In general, applications of hBN have utilized simple, flat flakes with thicknesses from a monolayer to hundreds of nanometers. Following lessons from silicon electronics, patterning hBN flakes is a potential strategy to enable more sophisticated functionalities, especially as the hBN typically occupies most of the device volume. In 2D electronics, its structure can influence active layers through modulation of their mechanical, electrostatic, and electromagnetic environments. In photonics, the hBN structure can lead to deterministic positioning of quantum emitters[32], waveguides[33], and metasurfaces[8]. However, in practice, designs are severely constrained by standard patterning techniques such as electron-beam lithography and focused ion-beam milling. If hBN flakes could be arbitrarily shaped to have freely varying profiles with features from tens of micrometers to tens of nanometers, previously unattainable structures for control of photons, electrons, and excitons would become possible.

To address this challenge, we have utilized thermal scanning-probe lithography due to its precise surface-structuring capabilities[19,20,34]. We spin-coat a thermally sensitive polymer resist over an hBN flake (Fig. 1a; Methods). A heated cantilever with a sharp tip is then raster-scanned across the film, removing polymer during its motion to form the desired freeform nanostructure in the resist. The resulting profile is subsequently transferred into the underlying hBN flake with reactive-ion etching.

The control offered by this approach allows the final structure to be designed using simple formulas. A grayscale bitmap controls the tip depth ($z$) at each in-plane pixel ($x, y$) during the scan. Thus, by converting mathematical profiles to high-resolution bitmaps, desired patterns can be easily fabricated. For example, Fig. 1 demonstrates an hBN flake structured with a portion of the Mandelbrot set. We chose this challenging fractal pattern as it requires features with continuously varying depth over a wide range of length scales. The design bitmap (Fig. 1b; Methods) assigns one of 256 depth levels (8 bit) to each 15×15 nm$^2$ pixel. Figure 1c presents an optical micrograph of the Mandelbrot pattern transferred into hBN (see also Extended Data Fig. 1). Colors arise in reflection due to thickness-dependent optical interference from the flake. However, because the pattern contains features beyond the resolution of our optical microscope, scanning-electron microscopy (SEM) is required to reveal the intricate self-similar features that persist down to tens of nanometers (Fig. 1d). Steps in $z$ of ~1 nm (3–4 atomic monolayers) are clearly visible in the SEM image as faint lines between terraces, consistent with the bitmap.

The limiting resolution in $x, y$ for such a pattern is set by several factors. Most importantly, the in-plane resolution decreases as the depth increases due to the conical shape of the probe[35,36]. A fresh probe tip has a radius of curvature down to ~3 nm and an estimated half-angle of 15–30°, which, combined with mechanical deformation in the resist, sets the minimum in-plane feature size for a given depth[35,36] (design rules are provided in Methods). Furthermore, the profile is initially written in the polymer film, which is ideally thin to avoid prolonged etching during the subsequent transfer step into hBN. However, if the polymer is too thin, unwanted thermal transport from the tip to the underlying substrate increases, limiting pattern quality. This trade-off sets a lower limit on the polymer film depth and, consequently, the pattern roughness that is accumulated during etching, which affects the minimum feature size in the final hBN structure.

Considering the above factors, we identified conditions for high-resolution patterning of hBN (Methods). We then designed and fabricated a 'freeform resolution target' that contains a controlled range of pattern depths and spatial frequencies. Specifically, as the spatial frequency increases the depth decreases due to the probe shape. The left half of Fig. 2a shows the bitmap (Methods); the right half presents atomic force microscopy (AFM) data of the experimental surface profile in hBN. The side-by-side comparison reveals good qualitative agreement. This is further supported by an SEM image of a full resolution target in hBN (Fig. 2b). To extract quantitative information, Fig. 2c compares the fabricated (AFM, blue circles) and targeted (red line) surface profiles for a single line scan through the middle of the resolution target (red dashed line, Fig. 2a). A root-mean-square (RMS) depth error of 1.3 nm was extracted. For comparison, the RMS roughness of our flat etched hBN was ~0.5 nm; the value for a pristine flake was reported[2] as ~0.1 nm. For the entire structure in Fig. 2a, we fit a 2D function to the full topography map, obtaining an RMS error of 3.5 nm (Extended Data Fig. 2). The fabricated profile follows the target well, even for shallow depth modulations (~2 nm amplitude) and increasing spatial frequencies (~150 nm period) at the edge of the pattern (Fig. 2c).

As this pattern had not yet reached the limit of our process, we extended the freeform resolution target in Fig. 2a at its corners to higher-resolution features at shallow modulation depths (~5 nm). Figure 2d shows an SEM image of the resulting hBN flake. This region spans periodicities from ~95 (bottom-left corner) to ~50 nm (top-right corner). Clear lattices persist over these length scales. Thus, to determine the ultimate resolution, we fabricated a series of structures, each with a fixed spatial frequency on a 2D square lattice (with periodicities from 35 to 25 nm). Figure 2e shows measured topography (AFM) for an hBN lattice with 29 nm periodicity and a depth of ~5 nm, which is close to the theoretical limit imposed by the probe geometry (Methods). This represented our limiting resolution for a high-quality pattern, based on a Fourier analysis of the topography data (Methods). Although smaller periodicities were possible to pattern (20 nm in the polymer; 25 nm in hBN), increasing disorder became apparent.

These results establish that we can create freeform nanostructures in hBN on optical and electronic length scales. We now demonstrate photonic microelements designed with simple mathematical expressions. Figure 3a shows an optical micrograph of an hBN flake with an array of phase plates, each defined by a continuous spiral height profile (Methods). Such patterns, which are unattainable with standard lithographic techniques, induce a twisted phase-front on transmitted light, producing optical orbital angular momentum[37] useful for controlling interactions between photons and electrons. Our structures are designed to impart a spiral $2\pi$ phase modulation on deep-ultraviolet photons at ~200 nm, where hBN shows lasing[1]. Figure 3b compares our bitmap (center) with the measured topography (AFM, outer region). By fitting the design function to the experimental profile, we obtain an RMS error of 4.0 nm (3.5%) (Methods; Extended Data Fig. 3). Such a low value for a continuously varying height profile confirms the fidelity of our approach. Furthermore, the array of phase plates arranged on the hBN flake verifies the potential for straightforward integration of multiple elements.

To characterize the optical quality of our photonic structures, we designed and fabricated a spherical hBN lens with a 100 µm radius of curvature (optical micrograph, Fig. 3c). The bitmap and the measured topography (AFM) are compared in Fig. 3d. By fitting a spherical function to the experimental profile, we extracted (Methods; Extended Data Fig. 4) a radius of curvature of 95 µm with an RMS error of 4.5 nm (2.8%). An impinging beam of collimated light should be focused by this structure. To observe this effect, we transferred the lens between two planar mirrors to form an optical microcavity[21] (Methods). Figure 3e depicts the cavity, which consists of a bottom distributed Bragg reflector (DBR), the hBN lens (blue), and a top moveable DBR. The red beam represents a stable cavity mode with transverse confinement due to the focusing of the lens. Figure 3f plots measured angle-resolved transmission spectra for our cavity (energy versus transverse wavevector, $k_x$), which contains longitudinal and higher-order transverse cavity modes. These are well described with a simple analytical model (Methods; Extended Data Fig. 5) in which the measured lens curvature and the reported refractive index of hBN[38] (2.12) were assumed, while the effective cavity length (33.7 µm) was used as a fitting parameter. The measured quality factor (Q) of the cavity resonances was ~8000 (Fig. 3f, inset), determined from the transmission spectrum at $k_x = 0$. These resonances represent a series of Ince-Gaussian modes[39] with specific transverse spatial profiles (calculated in Fig. 3g; observed in Fig. 3h). Thus, our results confirm high-quality optical microelements with properties useful for cavity quantum electrodynamic experiments with 2D materials[40]. Further, the ability to incorporate optical microstructures in hBN allows direct integration of optical functionalities into vdW heterostructures. For example, in Extended Data Fig. 6 sinusoidal gratings that couple photons into and out of guided modes in an hBN flake are demonstrated.

2D electronics can also benefit from hBN patterns with freeform profiles at shorter length scales. In particular, the propagation and interactions of electrons in nearby active layers can be manipulated. This possibility, known as electronic band-structure engineering, can be implemented through specific modulations of the hBN profile. While dielectric superlattices have recently been explored for this[14,16,23,24], the possible lattice structures were constrained to basic patterns (for example, arrays of holes) approachable by standard lithography. We lift these constraints by patterning hBN at nanoscale resolutions using mathematically defined freeform profiles. As a specific class of such structures, we introduce electronic Fourier surfaces[22], which superimpose a set of sinusoidal profiles to precisely control the spatial frequencies. Figure 4a shows a bitmap of a hexagonally symmetric electronic Fourier surface, defined by summing three sinusoids with 50 nm period, but rotated in plane by 0, 60, and 120° (Methods). The inset shows the fast Fourier transform (FFT) of the bitmap, revealing the hexagonal lattice symmetry. This pattern is then written in the polymer resist; Fig. 4b plots the topography measured during this process. From the real-space profile and FFT (inset), we see that the wavy hexagonal lattice is accurately reproduced. After reactive-ion etching, the same profile is replicated in hBN (Fig. 4c).

More importantly, such structures can be extended to more sophisticated profiles. Figure 4d shows a bitmap of an electronic Fourier surface with a moiré pattern, generated when two hexagonal lattices as in Fig. 4a (50 nm periodicity) are combined, with one rotated in-plane by 10°. Figure 4e,f confirms that this more complicated topography is reproduced in the polymer resist and hBN, respectively. The FFTs also clearly indicate the presence of two hexagonal lattices. Similarly, Fig. 4g–i demonstrates an electronic Fourier surface (the bitmap, polymer resist, and hBN flake, respectively) containing two aligned hexagonal lattices but with different periods (55 and 47 nm). Finally, for Fig. 4j–l, we designed and fabricated a quasicrystalline moiré profile from nine sinusoids (all with 50 nm periods, but 20° rotation between them). These structures confirm the ability to mathematically design and then experimentally generate high-resolution electronic lattices in hBN. Such profiles, which can cover large areas (10×10 µm$^2$, Extended Data Fig. 7), will modulate the electric field felt in a nearby active layer (Extended Data Fig. 8). Beyond electronic Fourier surfaces, our approach is amenable to any bitmap design within the limitations discussed above.

The high-fidelity structuring presented here exploits the simple combination of thermal scanning-probe lithography and reactive-ion etching to accurately replicate freely varying mathematical landscapes in hBN. In addition to integrated photonic microelements that modify photon flow, such control can modulate mechanical, electrostatic, and electromagnetic environments for 2D materials. For example, researchers are currently exploiting moiré periodicities induced by the rotation angle between two stacked monolayers of graphene (twisted bilayers)[11]. Freely patterned hBN should provide a more flexible and integrated approach to engineer strain, electronic band-structure, and cavity quantum electrodynamics. Thus, combining freeform hBN flakes with other 2D materials could provide a platform to access, discover, and exploit exotic states of matter in quantum materials.

**Online content** Methods, along with any Extended Data display items, are available in the online version of the paper; references unique to these sections appear only in the online paper.

**Acknowledgements** We thank S. Bisig, A. Imamoglu, R. Khelifa, N. Kiper, F. Könemann, M. Kroner, N. Maaskant, A. Popert, and L.B. Tan for helpful discussions and D. Dávila Pineda, U. Drechsler, and A. Olziersky from the Binnig and Rohrer Nanotechnology Center (BRNC) for technical assistance. We are also grateful to T. Taniguchi and K. Watanabe for providing the hBN crystal used in Fig. 3a. This project was funded primarily by ETH Zurich. D.T. and P.A.M. acknowledge support from the Swiss National Science Foundation under grant 200021-178909/1 and the European Union's Horizon 2020 program under Marie Sklodowska-Curiegrant MSCA-IF-OptoTransport (843842), respectively.

**Author contributions** N.L., D.T., and D.J.N. conceived the project. N.L. designed the structures with input from D.T., T.C., P.A.M., D.P., and D.J.N. N.L. and D.T. exfoliated the hBN flakes. N.L. patterned the polymer resist with thermal scanning-probe lithography and transferred the patterns to hBN with reactive-ion etching. N.L. characterized the flakes with optical and electron microscopy. N.L. and A.W.K. performed AFM topography measurements. N.L. analyzed the topography data with input from A.W.K. N.L. and A.W.K.

addressed the resolution limits of thermal scanning-probe lithography for hBN. P.A.M and T.C. conceived the use of patterned hBN as intracavity optical elements with input from N.L. D.T. transferred the hBN lens to the DBR substrate. T.C. performed the optical cavity measurements. T.C. and N.L. analyzed the optical cavity data. N.L. and D.P. performed optical measurements on the grating couplers. D.P. and N.L. analyzed the optical data from the grating couplers. D.T. performed electrostatic simulations with input from N.L. and D.J.N. N.L. and D.J.N. wrote the manuscript with input from all authors. D.J.N. supervised the project.

**Competing interests** The authors declare the following potential competing financial interests: N.L., D.T., and D.J.N. are seeking patent protection for ideas in this work. Readers are welcome to comment on the online version of the paper. Correspondence and requests for materials should be addressed to D.J.N. (dnorris@ethz.ch).

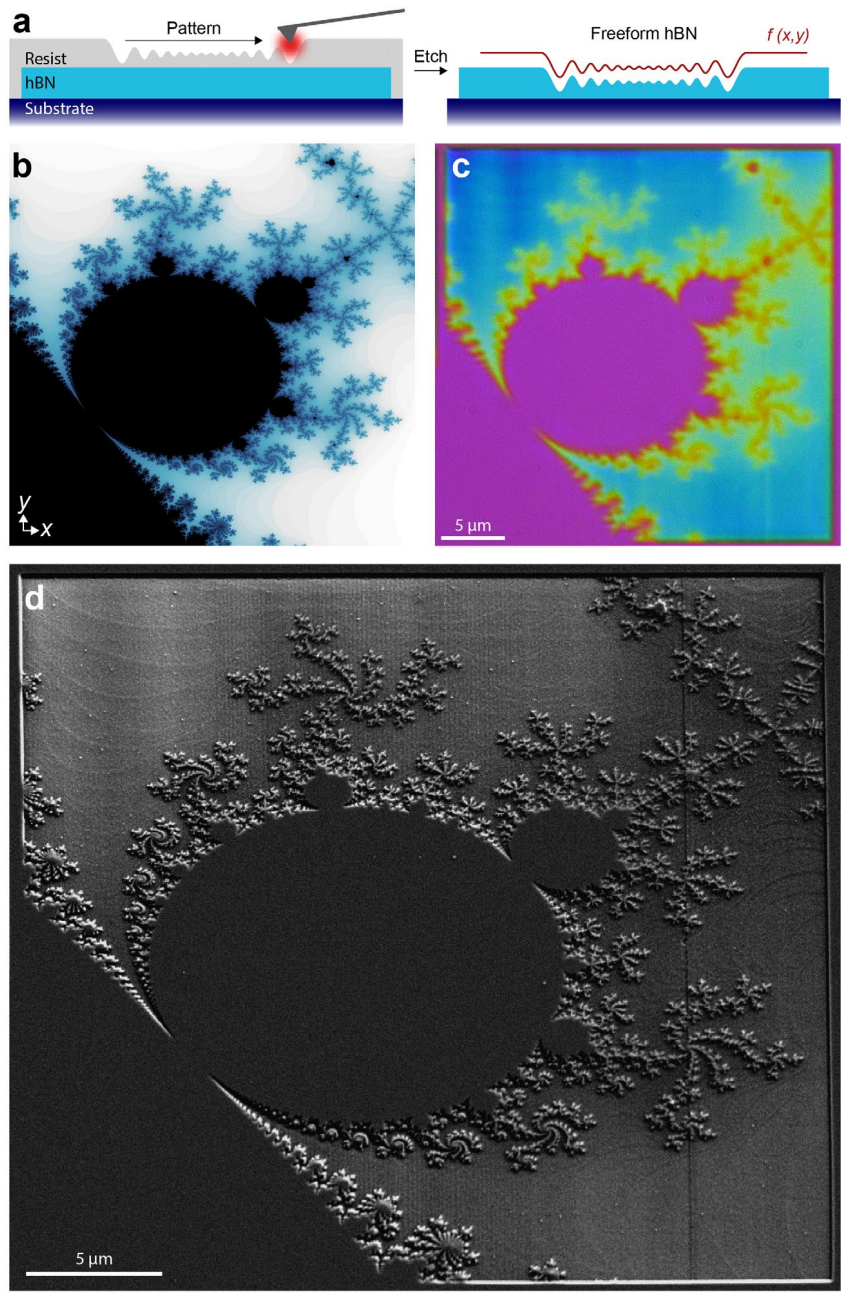

**Figure 1 | Freeform nanostructuring of hexagonal boron nitride (hBN). a**, Process flow showing the patterning (left) of a polymer resist using thermal scanning-probe lithography and the pattern transfer (right) from the polymer layer via reactive-ion etching into the hBN flake underneath. The resulting structure is a precise replica of the mathematical design, $f(x,y)$, represented by the red line. **b**, Bitmap utilizing a portion of the Mandelbrot set (see Methods). The color scale represents the height profile of the pattern when transferred to the polymer resist over a depth range of ~80 nm, with black (white) being the highest (lowest) point in the pattern. **c**, Optical microscope image of an hBN flake containing the Mandelbrot pattern in **b**, using the process in **a**. The colors arise in reflection due to thickness-dependent optical interference from the hBN flake. **d**, Scanning electron micrograph (SEM, 30° tilt) of the patterned hBN flake in **c**. The depth control in $z$ is directly visible as faint lines along the terraces in the surface (for example, in the upper left-hand corner of the image). The smallest steps visible in this image are ~1 nm in height, consistent with the bitmap, and correspond to 3–4 atomic monolayers.

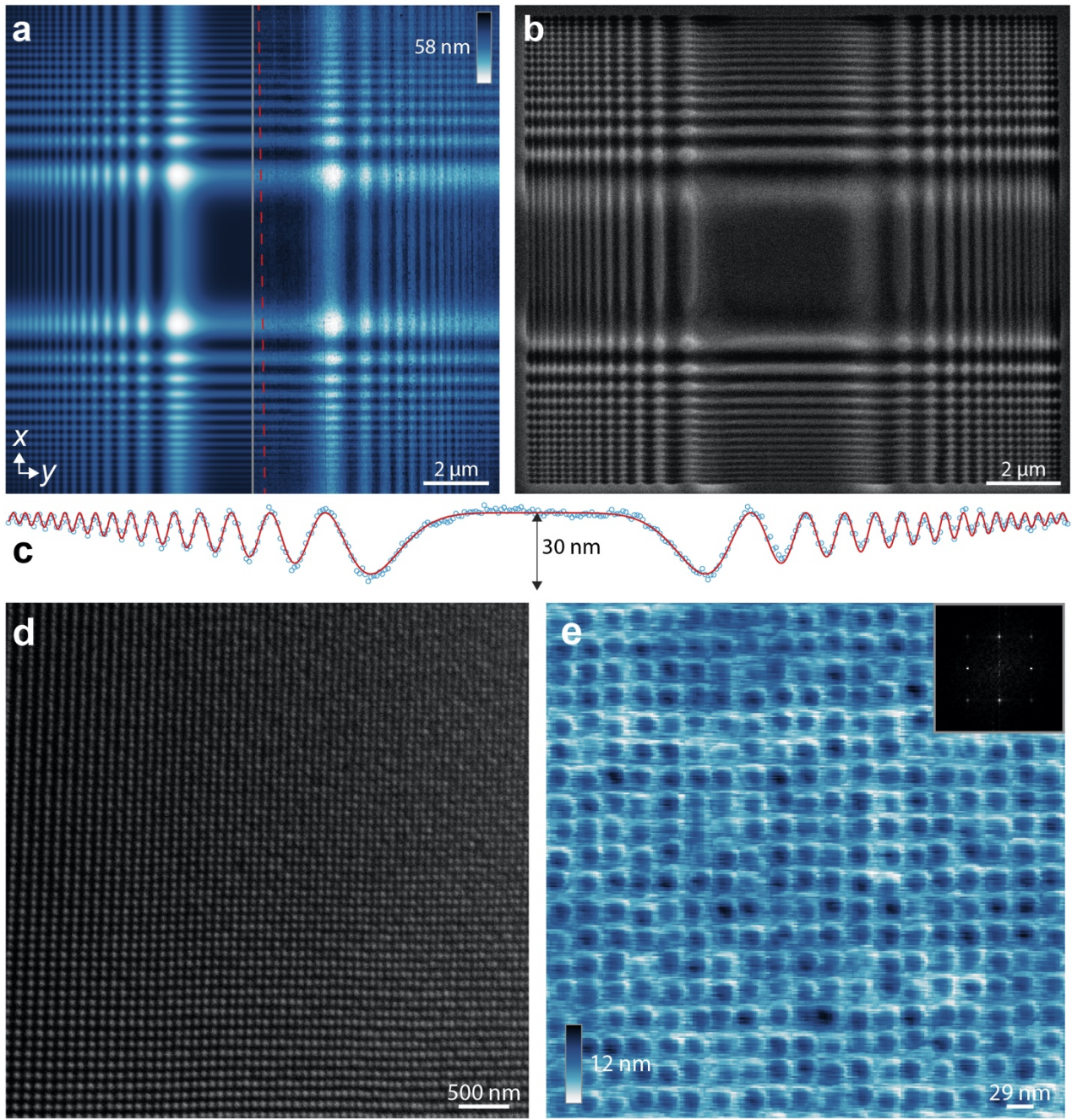

**Figure 2 | Capabilities of freeform nanostructuring of hBN. a**, Bitmap (left half) and measured atomic force microscope (AFM) data (right half) of a freeform resolution target in hBN. **b**, SEM (30° tilt) of a different hBN flake patterned with both sides of the resolution target in **a**. **c**, Measured (AFM, blue circles) and targeted (red line) surface topographies for the structure on the right half of **a**. The scan length is 14.61 µm long and represents a single line in the structure (red dashed line in **a**). The target function takes into account a slight distance miscalibration present in the thermal scanning probe. **d**, SEM (30° tilt) of the resolution target in hBN. A corner of the pattern in **a** was extended to higher spatial frequencies: from ~95 (bottom-left corner of the image) to ~50 nm (top-right corner). **e**, Measured topography (AFM) for a high-resolution square-lattice structure in hBN with a period of 29 nm, representing our resolution limit for patterning high-quality lattices (see Methods). The inset represents the fast Fourier transform (FFT) of the topography data. For all structural design parameters, see Extended Data Table 1.

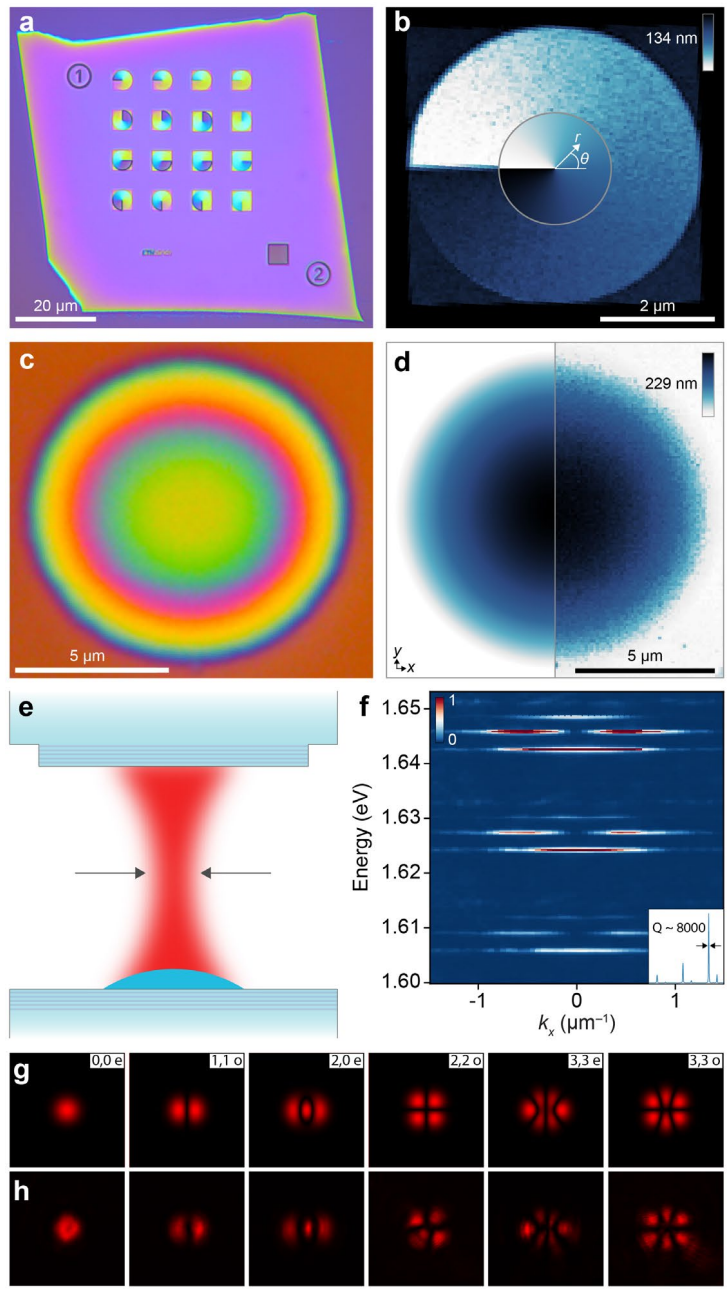

**Figure 3 | hBN optical microelements. a**, Optical micrograph of a patterned hBN flake (~130 nm thick) containing an array of optical microelements (spiral phase plates). **b**, Bitmap (central region) and measured topography (AFM, outer region) of a single hBN spiral phase plate from the flake in **a**. **c**, Optical micrograph of an hBN microlens with a spherical profile. **d**, Bitmap (left half) and measured topography (AFM, right half) of the hBN lens in **c**. **e**, Cartoon of the optical microcavity: a top mirror, bottom mirror, hBN microlens (blue), and cavity mode (red) with transverse confinement (black arrows). **f**, Angle-resolved transmission spectra (energy versus transverse wavevector, $k_x$) of the microcavity with hBN lens. The color scale represents normalized intensity. Horizontal streaks at fixed energy represent cavity modes. The inset shows a vertical line cut through the cavity spectra at $k_x = 0$, revealing a cavity quality factor, $Q$, of ~8000. **g**, Calculated transverse Ince-Gaussian mode profiles for our cavity geometry, labelled with their mode numbers $p$ and $m$, with e and o designating even and odd modes (Methods and ref. [39]). **h**, Measured transverse Ince-Gaussian mode profiles. For all structural design parameters, see Extended Data Table 1.

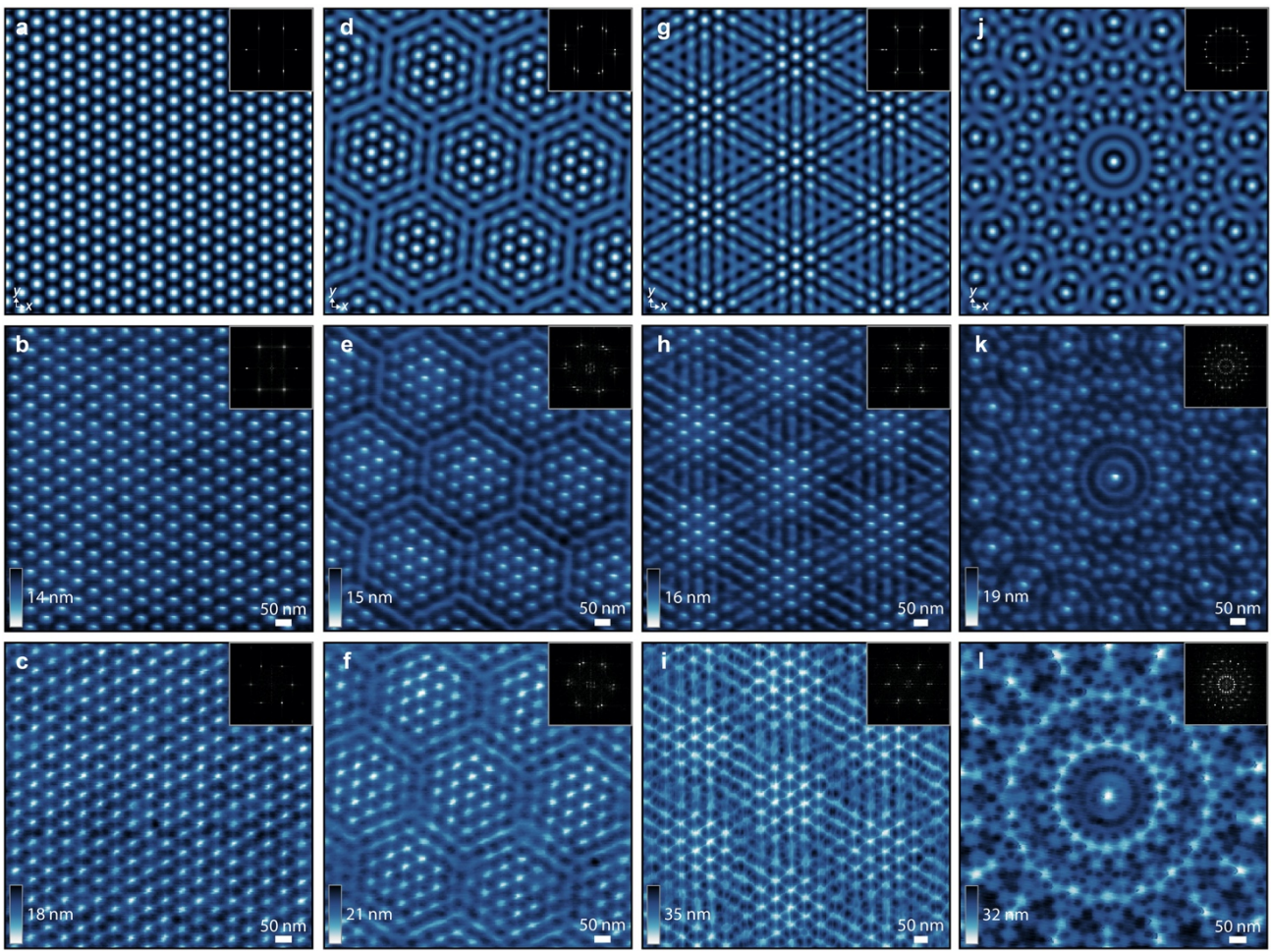

**Figure 4 | Electronic Fourier surfaces in hBN. a**, Bitmap of an electronic Fourier surface with a hexagonal lattice, defined by superimposing three sinusoids of period 50 nm rotated in plane by 0, 60, and 120°. **d**, Bitmap generated by superimposing two hexagonal lattices (as in **a**) with one lattice rotated in plane by 10°. **g**, Bitmap of two superimposed hexagonal lattices (as in **a**), with spatial periods of 55 and 47 nm and no in-plane rotation. **j**, Bitmap of nine superimposed sinusoids of period 50 nm, rotated in plane by 0, 20, 40, 60, 80, 100, 120, 140, and 160°. **b,e,h,k**, Measured topographies (obtained during patterning) of the bitmaps in **a,d,g,j**, respectively, written in the polymer resist. **c,f,i,l**, Measured topographies (AFM) of the patterns in **b,e,h,k**, respectively, etched into hBN flakes. Insets show FFTs for each pattern. For all structural design parameters, see Extended Data Table 1.

## Methods

**Freeform pattern design.** All patterns were designed using simple mathematical expressions. The height profile of the surface is represented in Cartesian coordinates for one dimension (1D) as $f(x)$ and in two dimensions (2D) as $f(x,y)$. The spiral phase plates are represented in polar coordinates as $f(r,\theta)$. The origin is located in the middle of the pattern, and the height profiles are defined relative to the unpatterned polymer surface, which is in the $xy$ plane at $z=0$. $z$ is orthogonal to the polymer surface with $+z$ pointing away from the substrate.

The Mandelbrot set used in Fig. 1 was calculated by iterating the nonlinear equation:

$$z_{n+1} = z_n^2 + c \tag{1}$$

where $z_n$, $z_{n+1}$, and $c$ are complex numbers. This equation can be used to generate a 2D image in which each pixel corresponds to a unique value of the complex number $c$, which is then plugged into the iterative equation with a starting value of $z_0 = 0$. The equation is then used to determine $z_1$, which is used to determine $z_2$, *etc*. The image of the entire Mandelbot set in Extended Data Fig. 1a was calculated using 500 iterations, where the $x$-axis, $\mathrm{Im}\{c\}$, ranges from 1.3 (left) to −1.3 (right), and the $y$-axis, $\mathrm{Re}\{c\}$, ranges from 1 (bottom) to −2 (top). The image was mapped onto a 1000×1000 pixel grid. The color scale outside the middle black region denotes the natural logarithm of the number of iterations required for the magnitude of $z_{n+1}$ to be greater than or equal to 2:

$$f(x,y) = \ln\left(n_{c_{x,y},|z_{n+1}|\geq 2}\right) \tag{2}$$

where deep blue (white) is a larger (smaller) number. The black region inside the boundary represents solutions that have a complex magnitude less than 2 after 500 iterations. When the number of iterations is taken to infinity, the values inside the boundary correspond to the Mandelbrot set.

The bitmap in Fig. 1b represents a portion of the Mandelbrot set (see red box in Extended Data Fig. 1a). It was calculated using the same procedure as above, except the center of the image is now located at –0.5135 – 0.5765$i$, and the $x$-axis, $\text{Im}\{c\}$, ranges from –0.501 (left) to –0.652 (right), and the $y$-axis, $\text{Re}\{c\}$, ranges from –0.4264 (bottom) to –0.6006 (top). This portion of the Mandelbrot set is mapped onto a 2000×2000 pixel grid and corresponds to 100 iterations. Finally, the bitmap is used to control the thermal scanning probe, where the depth of the pattern was chosen to range over approximately 80 nm, from 25 to 105 nm deep. The lateral size of the pattern was chosen to be 30×30 µm² with 15×15 nm² pixels.

The freeform resolution target in Fig. 2a was calculated with the expression:

$$f(x,y) = -(A - m|x|)\sin[(kx)^3]^2 - (A - m|y|)\sin[(ky)^3]^2 - \Delta \qquad (3)$$

where $A$ is the amplitude at the origin, $m$ is the slope that describes the linearly decreasing amplitude away from the origin, $k = 2\pi/\Lambda$ is the spatial frequency at the origin with $\Lambda = 12.5$ µm, and $\Delta$ is the vertical offset. The lateral size of the pattern was chosen to be 15.03×8.49 µm², mapped onto a 10×10 nm² pixel grid.

The spiral phase plates in Fig. 3a,b were calculated using:

$$f(r,\theta) = -m\theta - \Delta \qquad (4)$$

where $m$ is the slope describing the linearly decreasing height of the spiral phase plate as a function of the polar angle $\theta$, and $\Delta$ is the vertical offset. The lateral size of an individual phase plate was chosen to be 5×5 µm², mapped onto a 10×10 nm² pixel grid.

The spherical lens in Fig. 3c,d was calculated with:

$$f(x,y) = \sqrt{R^2 - x^2 - y^2} - \Delta \qquad (5)$$

where $R$ is the radius of curvature of the lens, chosen to be 100 µm, and $\Delta$ is the vertical offset. The lateral size of the pattern was chosen to be 20.02×20.02 µm², mapped onto a 20×20 nm² pixel grid.

The high-resolution pattern in Fig. 2e, the electronic Fourier surfaces in Fig. 4, and the large-area pattern in Extended Data Fig. 7 were calculated using:

$$f(x,y) = \sum_i A_i \cos(k_i[x\cos\theta_i + y\sin\theta_i]) - \Delta \qquad (6)$$

where $A_i$, $k_i$, and $\theta_i$ correspond to the amplitude, spatial frequency, and in-plane rotation angle, respectively, for component $i$. $\Delta$ is the vertical offset. For the high-resolution pattern in Fig. 2e, the lateral size was 580×580 nm², mapped onto a 2.9×2.9 nm² pixel grid. For the electronic Fourier surfaces in Fig. 4, the lateral size was 1×1 µm², mapped onto a 5×5 nm² pixel grid. For the large-area electronic Fourier surface in Extended Data Fig. 7, the lateral size was 10×10 µm², mapped onto a 5×5 nm² pixel grid.

The photonic grating couplers in Extended Data Fig. 6 were calculated using:

$$f(x) = A\cos(kx) - \Delta \tag{7}$$

where $A$ and $k$ correspond to the amplitude and spatial frequency, respectively. $\Delta$ is the vertical offset. The lateral size was 14×14 µm², mapped onto a 10×10 nm² pixel grid. For the parameters used in all formulas in this section, see Extended Data Table 1.

**Bitmap generation.** The mathematical expressions were converted into bitmaps, where the overall dimensions for the structure were chosen and the pattern was mapped onto a pixel grid (see above). The normalized depth of the pattern in the $z$-direction was discretized into 256 levels, corresponding to 8-bit precision. The physical depth of the patterns was assigned during patterning with the thermal scanning probe, where the total pattern depth was taken as an input, and the thermal scanning-probe software mapped the total depth onto the 8-bit depth levels.

**Materials.** Large-size bulk hBN crystals used in this work were purchased from 2DSemiconductors Inc., except the flake in Fig. 3a, which was provided by T. Taniguchi and K. Watanabe. Silicon wafers with 285 nm of SiO₂ were purchased from Alineason Materials Technology GmbH. For mechanical exfoliation of hBN we used adhesive tape purchased from Ultron Systems Inc. and 3M. The polymer resist for thermal scanning-probe lithography, PMMA/MA [AR-P 617, poly(methyl methacrylate-co-methacrylic acid), 33% copolymer, diluted to either 1.5 or 3 wt% in 1-methoxy-2-propanol] was purchased from Allresist GmbH. Silicon cantilevers (MBS 2019-12) for thermal-scanning probe lithography were bought from Heidelberg Instruments Nano. Acetone and isopropyl alcohol (IPA) were provided by the Binnig and Rohrer Nanotechnology Center (BRNC) at IBM Zurich, where the fabrication was performed.

**Sample fabrication.** Silicon wafers with 285 nm of $SiO_2$ were diced into chips of approximately 2×2 cm$^2$ area. Next, hBN flakes were deposited on the substrate using a mechanical exfoliation technique[41,42]. Briefly, repeated sticking and peeling of the hBN crystal to adhesive tape (Ultron Systems) thinned the bulk crystal down to flakes with thicknesses in the range of tens to hundreds of nanometers. After approximately 5–7 iterations of crystal thinning, Scotch-brand adhesive tape (3M) was applied to the crystal and removed. This tape with the thinned-down crystal was then placed on the Si/SiO$_2$ substrate (sticky side down) and a small vial filled with water was used as a weight to apply pressure to the top side of the tape. The substrate with tape (without weight) was moved onto a metal support, which was transferred to a hotplate, where it was heated for 1 min at 150 °C. After 1 min, the metal support containing the sample was removed and allowed to cool for 1 min. The tape was peeled off at a slow and uniform rate, leaving hBN flakes on the substrate ready for patterning.

PMMA/MA polymer resist was spin-coated onto the chip containing hBN flakes using a two-step procedure. The resist was dispensed onto the sample surface, which was then accelerated at 500 r.p.m. s$^{-1}$ to 500 r.p.m. for 5 s. Next, the spinning speed was increased at 2000 r.p.m. s$^{-1}$ to the final spin speed (2000-6000 r.p.m., depending on the required thickness) for a total time of 40 s. After spin-coating, the sample was baked at 180 °C for 5 min.

The sample was then placed on the stage of the thermal scanning-probe lithography tool (NanoFrazor Explore, Heidelberg Instruments Nano); the flake of interest was centered and rotationally aligned under the optical microscope of the tool. A cantilever was loaded into the cantilever holder, which was then attached to the Nanofrazor scan head. The tip was brought near the sample surface, and an automated approach function was used to find the sample surface and bring the tip into contact. For all freeform patterns other than the high-resolution structures shown in Fig. 2e and Fig. 4, the tip was then moved away from the flake to perform calibration scans. After calibration, the tip was optically aligned over the flake of interest. Next, the thermal scanning probe performed a topography scan of the polymer surface on top of the flake for fine alignment of the pattern and to ensure that the surface was relatively flat and smooth in the local pattern area. The thermal scanning probe was set to an initial temperature of 950 °C, and it then started fabricating the desired pattern, allowing the feedback to adjust the patterning conditions until the fabricated pattern matched the design pattern. The scan proceeded until the entire pattern was written in the polymer resist. Afterwards, the thermal scanning probe was available to create the next pattern on either the same flake or a different flake on the same chip.

For the high-resolution patterns, a fresh cantilever was loaded as in the procedure above, however, no calibration scans were performed. This minimized the contamination that builds up on the tip during patterning, which limits the resolution. The fresh cantilever was positioned directly over the flake of interest, and the patterning was initiated, but in this case the feedback was turned off. This was done to have maximum control over the patterning conditions. The starting temperature was set between 950–1000 °C, and a minimal value of writing force was applied between the tip and substrate to observe conditions where no pattern was generated. The onset of patterning was initiated by repeatedly writing the same structure, where the write force was slowly increased for each scan until the desired pattern was observed in the polymer resist. Further adjustments to the temperature and write force were implemented manually and iteratively, until optimal conditions were found for the high-resolution pattern of interest. Once these conditions were identified, the patterns of interest were consecutively written in the polymer film. This typically resulted in a few tens of patterns before the cantilever had to be changed due to contamination build-up on the tip.

Once thermal scanning-probe lithography was completed, the pattern was transferred to the underlying hBN flake via inductively coupled plasma (ICP) etching[43] (Oxford Instruments, PlasmaPro) using a gas content of 50 sccm $SF_6$. The etching was performed with a chamber pressure of 40 mTorr, a forward power of 75 W, and at a rate of ~2 nm s$^{-1}$ until the polymer resist was removed. The pattern was transferred to the underlying hBN with approximately 1:1 depth, indicating little to no pattern amplification. After etching, the sample was sonicated for 2 min in acetone, rinsed with IPA, and blown dry with $N_2$ gas.

**Design rules for in-plane resolution versus pattern depth.** The conical shape of the thermal scanning probe, combined with mechanical deformations in the polymer resist, set the lower limit on in-plane periodicity for a given depth. This limit can be estimated as follows. A fresh probe has a tip diameter at the apex as low as 6 nm and half-angle of 15–30°. We note that these quantities vary from probe to probe due to fabrication tolerances. Thus, the probe width is a function of the distance from the tip, set by the pattern depth. For a periodic structure, the relationship between the minimum periodicity, $\Lambda_{min}$, and the pattern depth, $d$, can be written as:

$$\Lambda_{min} = 2w(d) = 2(w_0 + 2d \tan \theta_{\text{half}} + w_m) \qquad (8)$$

where $w(d)$ is the width of the indent, $w_0$ is the probe width at the apex, $\theta_{\text{half}}$ is the opening half-angle of the probe tip, and $w_m$ represents additional feature broadening beyond the probe shape due to mechanical deformations. A detailed discussion of this topic is available[36]. The prefactor of 2 arises from the assumption that a periodic structure will have a period twice as wide as the indent caused by the probe. We note that in practice contamination will increase the size of the probe, which can further increase $\Lambda_{min}$. Furthermore, roughness accumulated during etching can additionally increase $\Lambda_{min}$ for the final pattern in hBN.

**Surface-topography characterization.** The topography of the patterns in the polymer resist was measured by the thermal scanning probe during the writing process. The final patterns in hBN were measured with an AFM (Bruker, Dimension FastScan, NCHV-A cantilever) using tapping mode in ambient conditions. The topography data was processed via a custom MATLAB script that performed row alignment, plane levelling, and function fitting to extract structural parameters, RMS roughness, and error values.

The high-resolution (25–35 nm periodicity) square lattices were measured using an AFM (Veeco Dimension V with Nanosensors PPP-NCHR probes) in non-contact mode. To extract a quantitative measure of the high-resolution lattice quality, Fourier analysis was used on the measured topography data. The 2D FFT of the topography data revealed prominent peaks (along $k_x$ at $k_y = 0$, and along $k_y$ at $k_x = 0$) that correspond to the fundamental spatial frequency of the lattice. The ratio of the fundamental peak height to the next-highest peak in the Fourier spectrum was taken as a quantitative metric for the lattice quality. We chose a threshold of 5 for this ratio as our criterion for a high-quality lattice. The lattice with 29 nm periodicity was the shortest period that had a ratio greater than 5 (5.05). Thus, 29 nm was taken as our limiting spatial resolution.

**Transfer of hBN lens.** After etching and topography characterization, the hBN lens was transferred to the DBR substrate using a standard dry polymer transfer method[44] under inert atmosphere in a glove box.

**Optical cavity measurements.** After the hBN lens was transferred to the first DBR substrate, another moveable DBR was brought close to the first (distance of ~33 μm), forming an optical microcavity with two planar mirrors. The cavity modes were excited with a broadband source (Fianium supercontinuum laser, NKT Photonics) at a range of incident angles, where the transmitted light was collected and the Fourier plane was spectrally dispersed onto a liquid-nitrogen-cooled charge-coupled-device (CCD) camera (Extended Data Fig. 5a). The control cavity spectra for two planar mirrors without the lens was taken by passing the excitation beam through the flat part of the hBN flake (Extended Data Fig. 5b). In this case, unstable, low-Q longitudinal modes are observed. The cavity spectra for the hBN lens was taken when the excitation beam was passed through the lens, resulting in the cavity transmission spectra shown in Fig. 3f and Extended Data Fig. 5c. To measure the transverse intensity profiles of the different cavity modes, the sample was excited with a narrowband laser (Solstis, M Squared) which is tuned to the energy of the cavity mode of interest. The transverse mode intensity distribution is then recorded on a CCD camera (Chameleon3, FLIR Systems), by imaging the surface of the sample with a microscope with 15× magnification.

**Modeling of cavity modes.** The frequencies of the different transverse and longitudinal modes of the microlens cavity are given by the resonant standing-wave condition[45]

$$\omega_{qmn} = \frac{c}{2*L_{\text{eff}}}\left\{q + (n+m+1)\frac{\cos^{-1}[\pm\sqrt{g_1 g_2}]}{\pi}\right\} \tag{9}$$

where {q, n, m} is a set of integers labelling the longitudinal and transverse mode orders, respectively, $L_{\text{eff}}$ is the effective cavity length, $g_1 = 1$ for the flat DBR mirror, and

$$g_2 = 1 - (n_{\text{hBN}} - 1) * L_{\text{eff}}/r_{\text{hBN}} \tag{10}$$

for the DBR mirror with the hBN lens. The refractive index of hBN, $n_{\text{hBN}}$, was set to 2.12 (ref. [38]), and the radius of curvature of the microlens was set to $r_{\text{hBN}} = 95$ μm, according to our AFM measurement. As shown in Extended Data Fig. 5, this simple analytical solution to the resonator mode equation reproduces the measured spectrum.

**Guided-mode coupling measurements.** The hBN flake, optical setup, and optical data used to characterize the in-plane guided-mode coupling to and from the hBN flake is depicted in Extended Data Fig. 6. Two aligned grating couplers, each with a measured period of 282 nm (Extended Data Fig. 6a), were patterned in an hBN flake of 255 nm thickness, where the two gratings were separated by ~20 μm. To observe in-plane guided-mode coupling, we performed measurements where one grating coupler was illuminated to couple light into a guided mode of the flake and the second coupler was used to couple light out of the guided mode. The output was then collected and imaged onto a spectrometer for dispersed *k*-space imaging[46].

The measurements were performed using an inverted optical microscope (Nikon, Eclipse Ti-U) with a 50× air objective [Nikon, TU Plan Fluor, numerical aperture (NA) of 0.8], imaging both grating couplers simultaneously. As a light source, we used a broadband halogen lamp. A small, circular aperture located in the real-space image plane before the objective ensured that only the first grating coupler was illuminated and that no light was incident on the second grating coupler. After the aperture, the light was reflected onto the sample using a beamsplitter, where it passed through the objective lens and was focused on the sample. Therefore, in this configuration, the first grating coupler was illuminated with broadband light under all possible incident angles (limited by the NA of the objective).

To observe guided-mode coupling, the outcoupled light at the second grating coupler was collected by the same objective, transmitted through the same beamsplitter and passed through another small circular aperture located in the real-space image plane. This second aperture was used to ensure that light was only collected from the second grating coupler on the hBN flake. The back focal plane of the microscope objective was then imaged onto the entrance slit of an imaging spectrograph (Andor Shamrock 303i) and captured by a digital camera (Andor Zyla PLUS sCMOS).

Dispersed *k*-space measurements[46] were performed by inserting a grating (150 lines mm$^{-1}$ blazed at 500 nm) into the imaging path of the spectrometer, such that the outcoupled light was spectrally dispersed along one axis of the camera pixel array. A slit was closed to a width of 100 μm along the $k_x$ axis at $k_y \approx 0$. Here, $k_x$ is the wavevector direction that corresponds to the modulated direction of the two grating couplers. Thus, the setup allowed for an angle- and wavelength-resolved measurement of the light coupled out at the second grating coupler with a single image. To eliminate the effects of background and stray light incident on the camera, a reference measurement on the flat portion of the same hBN flake was performed and subtracted. Furthermore, a linear polarizer was placed in the collection path to selectively measure s- or p-polarized light. We note that s-polarized light corresponds to a transverse-electric (TE) waveguide mode, and p-polarized light corresponds to a transverse-magnetic (TM) mode. A schematic of the optical setup is shown in Extended Data Fig. 6b.

Extended Data Fig. 6c shows a dispersed *k*-space measurement, where we observe two branches corresponding to outcoupled light that originates from two guided modes of the hBN flake. The lower branch corresponds to the $TE_0$ mode, and the upper branch corresponds to the $TE_1$ mode. We note that the broken inversion symmetry around $k_x = 0$ occurs because the light is coming from a single direction, starting at the first (incoupling) grating coupler and moving towards the second (outcoupling) grating coupler. We confirmed that if the light propagation direction was reversed, the data shown in Extended Data Fig. 6c flipped around $k_x = 0$. A comparison between measurements for s- and p-polarized light shows that the outcoupled light is s-polarized (TE). This is consistent with the theoretical expectation that, for this wavelength range and flake structure, only the $TE_0$ and $TE_1$ modes of the hBN flake are coupled (see Methods section 'Theoretical calculations of the guided modes').

**Theoretical calculations of the guided modes.** To compare the observed guided modes to theoretical predictions, we performed electromagnetic simulations (Lumerical). We considered a system with an hBN layer that is surrounded by an infinite half-space of air above and an infinite half-space of SiO₂ below. For the hBN layer we used the measured thickness of 255 nm and the previously measured in-plane and out-of-plane indices of refraction[38]. Calculating the eigenmodes for the experimental wavelength range and their corresponding effective indices, $n_{\text{eff}}(\lambda_0)$, yields the effective wavevector $k_{\text{eff}} = n_{\text{eff}}(2\pi/\lambda_0)$ of the in-coupled light propagating inside the waveguide as a function of the free-space wavelength $\lambda_0$. The outcoupler grating then couples light from the guided mode to light in free space with an in-plane momentum according to $k_{x,\text{out}} = k_{\text{eff}} - k_{\text{g}}$. Here, $k_{\text{g}} = \frac{2\pi}{\Lambda_{\text{g}}}$ is the wavevector associated with the period of the grating coupler with $\Lambda_{\text{g}}$ equal to 282 nm. Therefore, each guided mode in the measurements of Extended Data Fig. 6c will appear as a branch that follows the dispersion relation $E(k) = \hbar c k_{\text{eff}}$ but at an in-plane momentum shifted by $-k_{\text{g}}$ into the light cone. Here, $\hbar$ is the reduced Planck constant and $c$ is the speed of light in vacuum. Extended Data Fig. 6c shows the theoretical expectation of the $TE_0$ and $TE_1$ modes obtained from our numerical calculations. The shaded regions correspond to a ±3% uncertainty on the reported indices of refraction for hBN[38].

**Electrostatic simulations.** Electrostatic simulations of devices incorporating patterned hBN structures are performed by solving Poisson's equation with the finite-element solver COMSOL. The devices consisted of an upper and lower hBN flake that together encapsulated a 2D active layer in between (Extended Data Fig. 8a). The lower hBN flake has a thickness of 10 nm and is flat. The upper hBN flake was structured such that its top surface was described by the functional relations provided in Extended Data Table 1 with a depth modulation of 20 nm and a 5 nm spacing between the bottom of the pattern and the bottom surface of the flake. Furthermore, we assumed an isotropic dielectric constant for hBN with a value of 4.5 and considered a potential difference of 1 V between the top and bottom gates, represented by the top and bottom surfaces of the two hBN flakes, respectively.

**Data availability.** The data supporting the findings of this study are available from the corresponding author on reasonable request.

| Figure | Parameters | | | | | Height profile |
|---|---|---|---|---|---|---|
| | $A$ (nm) | $m$ (nm µm$^{-1}$) | $\Lambda$ (µm) | $s$ | $\Delta$ (nm) | $f(x,y)$ |
| Fig. 2a | 28.5 | 3.7 | 12.5 | 2 | 0 | $-(A-m|x|)\sin[(kx)^3]^s - (A-m|y|)\sin[(ky)^3]^s - \Delta$ |
| Fig. 2b | 42.8 | 5.5 | 12.5 | 2 | 0 | |
| Fig. 2d | 2.5 | 0 | 12.5 | 1 | 2.5 | |

| Figure | $i$ | $A_i$ (nm) | $\Lambda_i$ (nm) | $\theta_1$ (deg) | $\theta_2$ (deg) | $\theta_3$ (deg) | $\theta_4$ (deg) | $\theta_5$ (deg) | $\theta_6$ (deg) | $\theta_7$ (deg) | $\theta_8$ (deg) | $\theta_9$ (deg) | $\Delta$ (nm) | $f(x,y)$ |
|---|---|---|---|---|---|---|---|---|---|---|---|---|---|---|
| Fig. 2e | 1,2 | 1.9 | 29 | 0 | 90 | – | – | – | – | – | – | – | 3.8 | $\sum_i A_i \cos[k_i(x\cos\theta_i + y\sin\theta_i)] - \Delta$ |
| Fig. 4a | 1,2,3 | 2.2 | 50 | 0 | 60 | 120 | – | – | – | – | – | – | 6.7 | |
| Fig. 4d | 1–6 | 1.1 | 50 | 0 | 60 | 120 | 10 | 70 | 130 | – | – | – | 6.7 | |
| Fig. 4g | 1–6 | 1.1 | 47,55 | 0 | 60 | 120 | 0 | 60 | 120 | – | – | – | 6.7 | |
| Fig. 4j | 1–9 | 0.8 | 50 | 0 | 20 | 40 | 60 | 80 | 100 | 120 | 140 | 160 | 6.9 | |
| Ext. Data Fig. 7 | 1,2,3 | 2.2 | 50 | 0 | 60 | 120 | – | – | – | – | – | – | 6.7 | |

| Figure | $m$ (nm rad$^{-1}$) | $\Delta$ (nm) | $f(r,\theta)$ |
|---|---|---|---|
| Fig. 3b | 18.3 | 10.0 | $-m\theta - \Delta$ |

| Figure | $R$ (µm) | $\Delta$ (µm) | $f(x,y)$ |
|---|---|---|---|
| Fig. 3d | 100 | 100.0 | $\sqrt{R^2 - x^2 - y^2} - \Delta$ |

| Figure | $A$ (nm) | $\Lambda$ (nm) | $\Delta$ (nm) | $f(x)$ |
|---|---|---|---|---|
| Ext. Data Fig. 6 | 20.0 | 280 | 20.0 | $A\cos(kx) - \Delta$ |

**Extended Data Table 1 | Design parameters for freeform nanostructures.** Design parameters for freeform nanostructures based on analytical functions. The functions are defined for the pattern in the polymer surface, written by the thermal scanning probe, where $x$ and $y$ lie in the sample plane and $z$ is perpendicular, pointing away from the substrate. A right-handed coordinate system is used. The origin is placed in the middle of the pattern in the $x$ and $y$ directions, and at the unpatterned polymer surface in the $z$ direction. The height of the patterned surface is defined relative to the unpatterned polymer surface at $z = 0$. The parameters $A$ and $\Delta$ have been rounded to the nearest 0.1 nm, and $m$ has been rounded to the nearest 0.1 nm µm$^{-1}$ and 0.1 nm rad$^{-1}$. See Methods for further details.

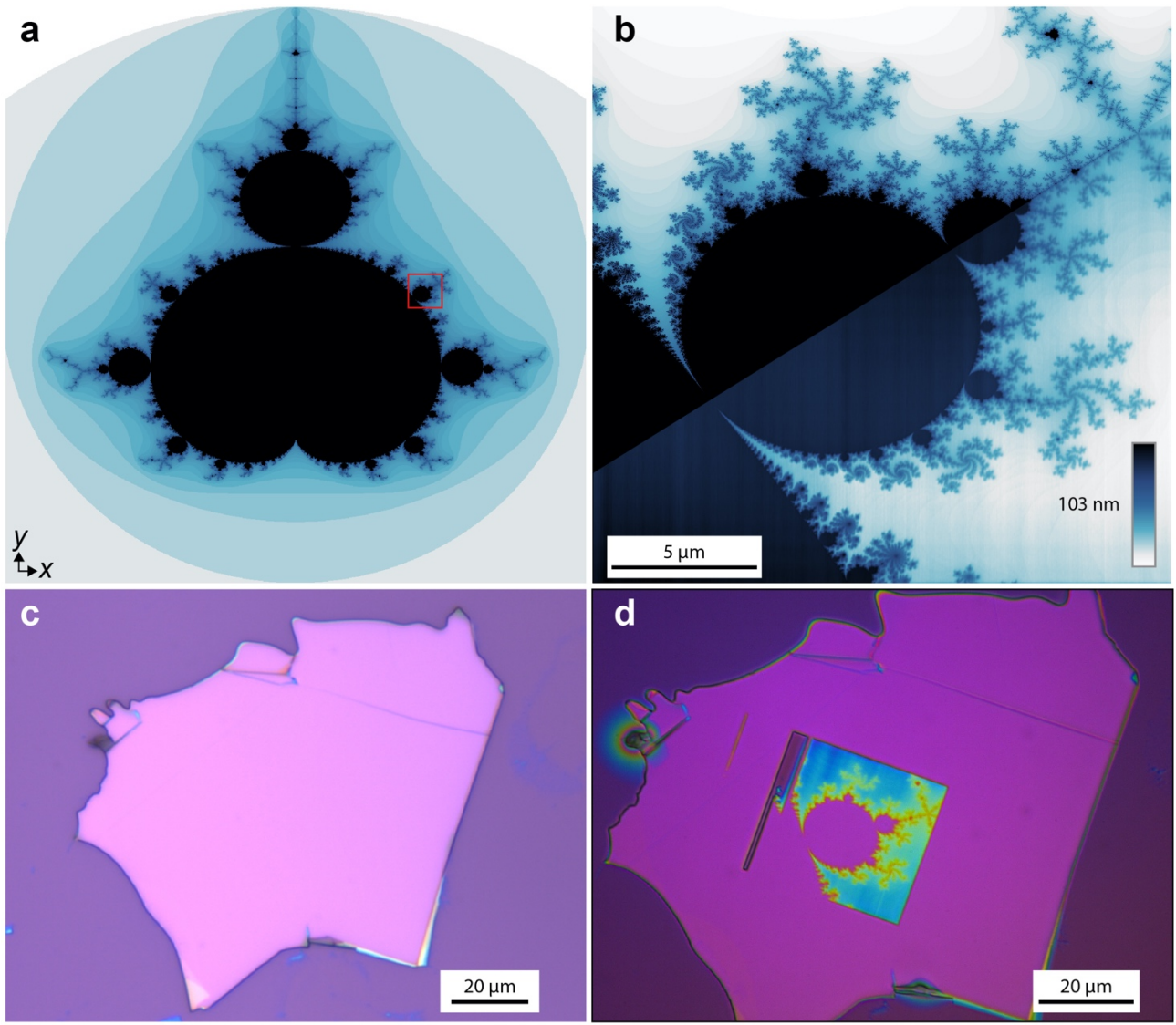

**Extended Data Figure 1 | Design and fabrication of a Mandelbrot pattern in an hBN flake. a**, Full Mandelbrot bitmap (see Methods). The red box indicates the portion of the pattern used in Fig. 1. **b**, Portion of the Mandelbrot set used in Fig. 1 (red box in **a**). Top left: bitmap. Bottom right: topography data of the polymer measured by the thermal scanning probe during patterning. **c**, Optical microscope image of the hBN flake (~126 nm thick) before the polymer resist was spun on top. **d**, Optical microscope image of the hBN flake from **c** patterned with a portion of the Mandelbrot set (red box in **a**). The image was taken after etching and cleaning (Methods), where all residual polymer is assumed to be removed.

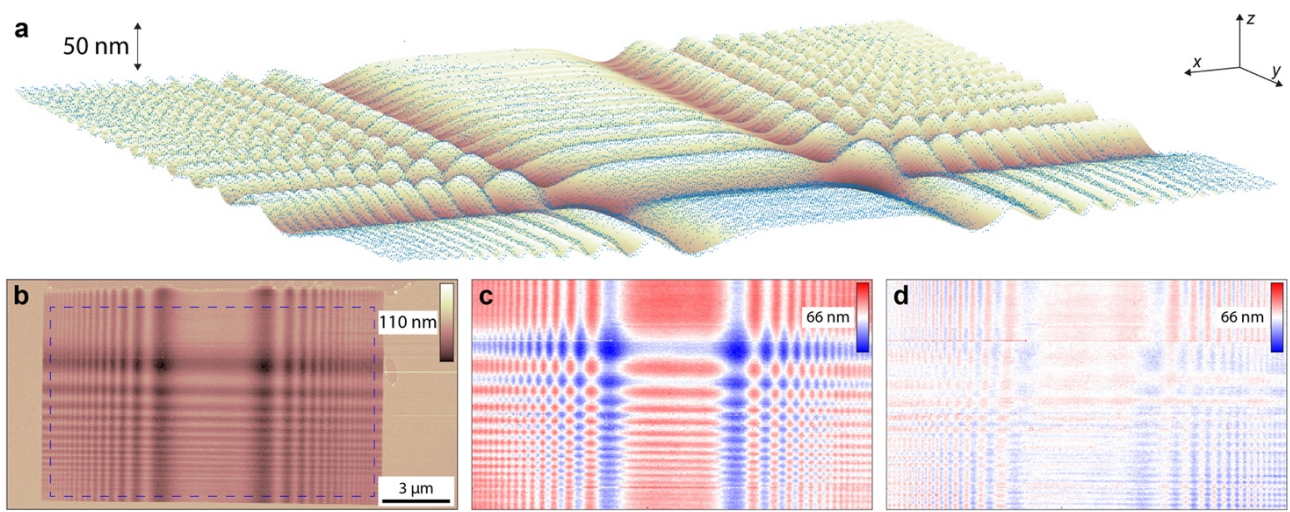

**Extended Data Figure 2 | Topography characterization of hBN freeform resolution target. a**, Measured topography (AFM, blue points) and fitted function (yellow/brown surface) of the freeform resolution target in hBN (Fig. 2a). **b**, Measured topography (AFM) of the freeform resolution target in hBN. The dashed blue box indicates the region used for fitting the 2D freeform function. **c**, Measured topography of the structure in **b**, plotted only for the fit region (blue box in **b**), scaled from the minimum depth value to the maximum depth value and centered at zero. **d**, Residual error between the data and fitted function, plotted for the fit region as in **c**. For comparison, the data is scaled over the same range as in **c**, centered at zero. For all structural design parameters, see Extended Data Table 1.

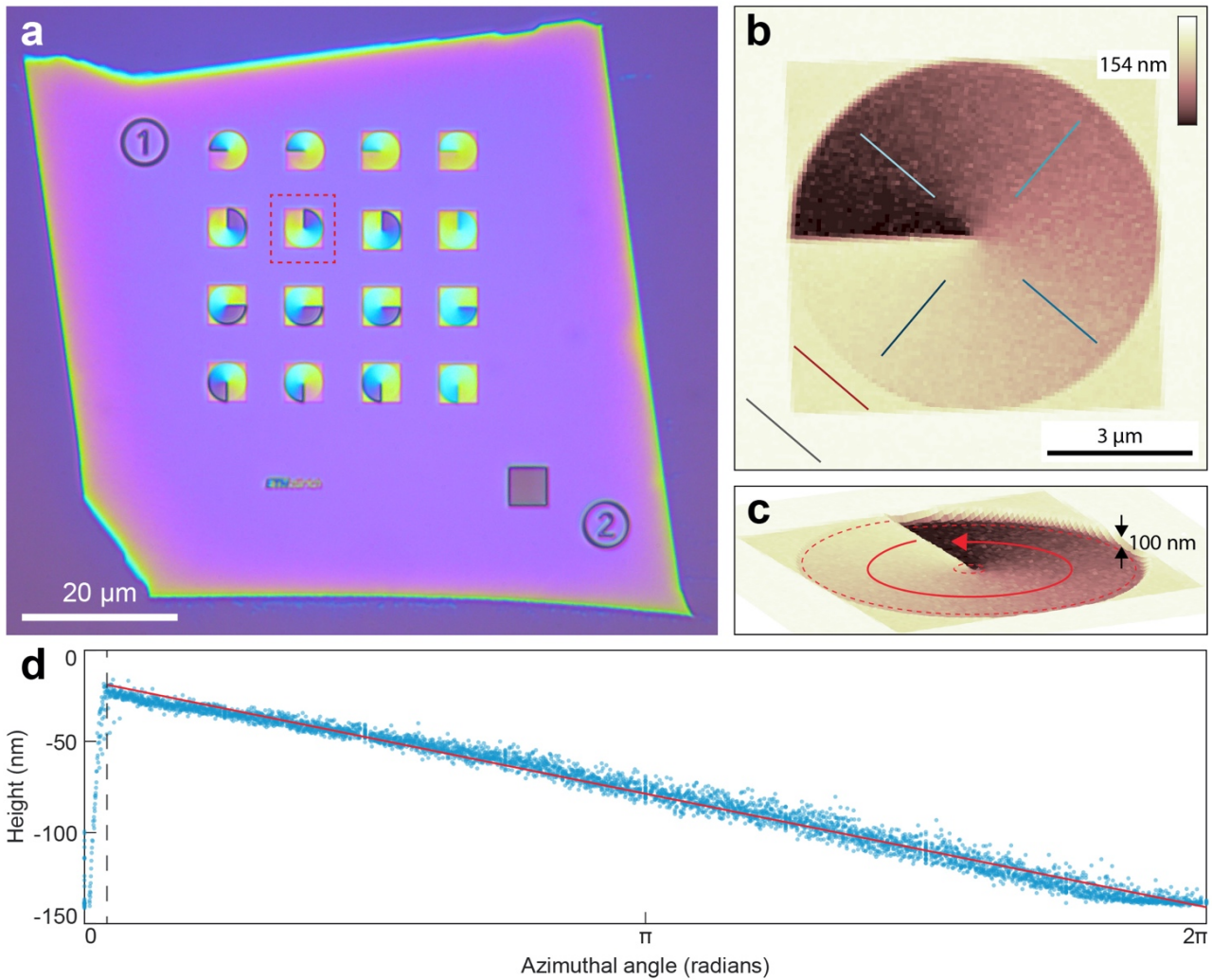

**Extended Data Figure 3 | Topography characterization of hBN spiral phase plate. a**, Optical microscope image of the hBN flake from Fig. 3a,b (~130 nm thick) containing an array of spiral phase plates. These structures will impart a spiral $2\pi$ phase modulation on deep-ultraviolet photons at ~200 nm. **b**, Measured topography (AFM) of the phase plate indicated with the dashed red box in **a**. The overlaid lines (grey, red, and series of blue) indicate line cuts used to determine the RMS roughness of the unpatterned flake (0.4 nm), of a flat region on the patterned flake (0.8 nm), and of the freeform pattern (1.7 nm), respectively. **c**, Same data as in **b**, shown as a three-dimensional topography plot. **d**, Measured topography (AFM, blue points, collected from the annular region between the dashed red lines in **c**) and fitted function (red line) of the spiral phase plate, expressed as a linear function in polar coordinates (Methods). The linear fit returns an RMS error of 4.0 nm for a pattern depth of 115 nm (3.5%). For all structural design parameters, see Extended Data Table 1.

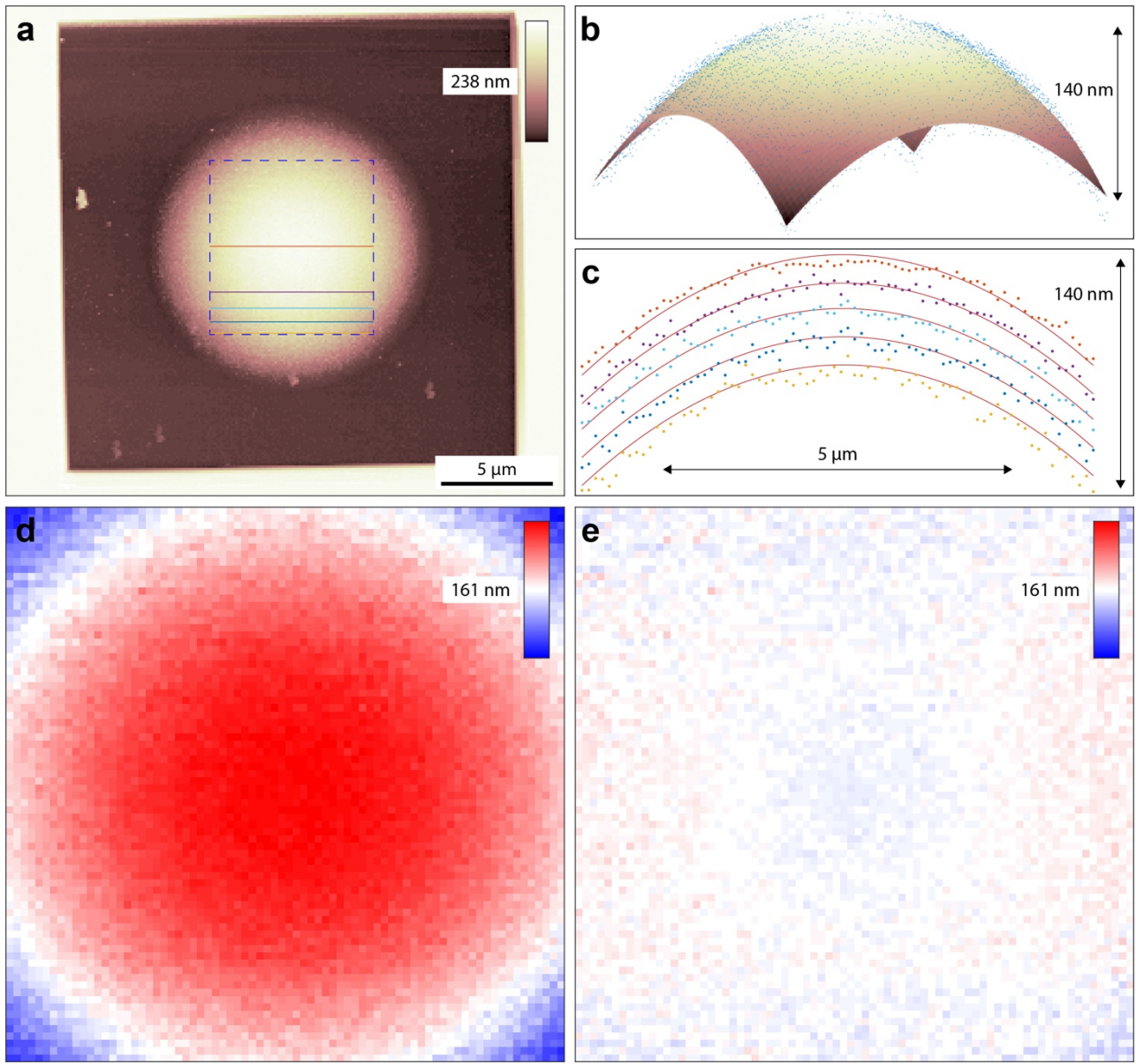

**Extended Data Figure 4 | Topography characterization of hBN microlens. a**, Measured topography (AFM) of the spherical hBN microlens in Fig. 3c,d. The dashed blue box indicates the region used for fitting the 2D spherical function. The colored horizontal lines indicate line cuts shown in panel **c**. **b**, Measured topography (AFM, blue points) and fitted function (yellow/brown surface) of the hBN microlens, resulting in an RMS error of 4.5 nm (2.8%). **c**, Line cuts (red) through the 2D fitted function, plotted with the corresponding line scans of the measured topography in **a** (points colored to match the horizontal lines in **a**). **d**, Measured topography of the structure in **a**, plotted only for the fit region (dashed blue box in **a**), scaled from the minimum depth value to the maximum depth value and centered at zero. **e**, Residual error between the data and fitted function, plotted for the fit region as in **d**. For comparison, the data is scaled over the same range as in **d**, centered at zero. For all structural design parameters, see Extended Data Table 1.

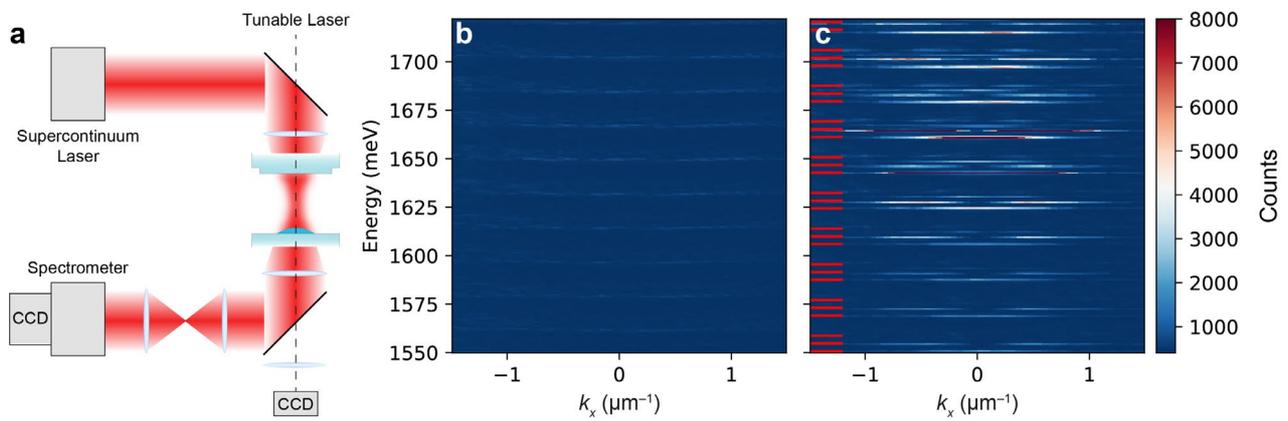

**Extended Data Figure 5 | Measurement and modeling of microcavity modes. a**, Diagram of measurement setup, showing the hBN microlens from Fig. 3c,d placed between two cavity mirrors. For the angle-resolved transmission spectra shown in **b,c**, and Fig. 3f, the cavity is fed with a broadband source (supercontinuum laser) at a range of incident angles. A 4$f$ imaging setup (where $f$ is the focal length of the optics) is used to measure angle-resolved cavity transmission, which is spectrally dispersed by a spectrometer and then imaged on a liquid-nitrogen-cooled CCD camera. For the measured transverse-mode profiles shown in Fig. 3h, a narrowband tunable laser was used as the source, and the transmitted light was imaged onto a CCD. **b**, Angle-resolved transmission spectra (energy versus in-plane wavevector, $k_x$) for a cavity with a flat hBN flake, revealing unstable cavity modes supported by the two planar mirrors. **c**, Angle-resolved transmission spectra (as in **b**) for a cavity that includes the hBN microlens. The red lines (left) indicate calculated mode energies (Methods).

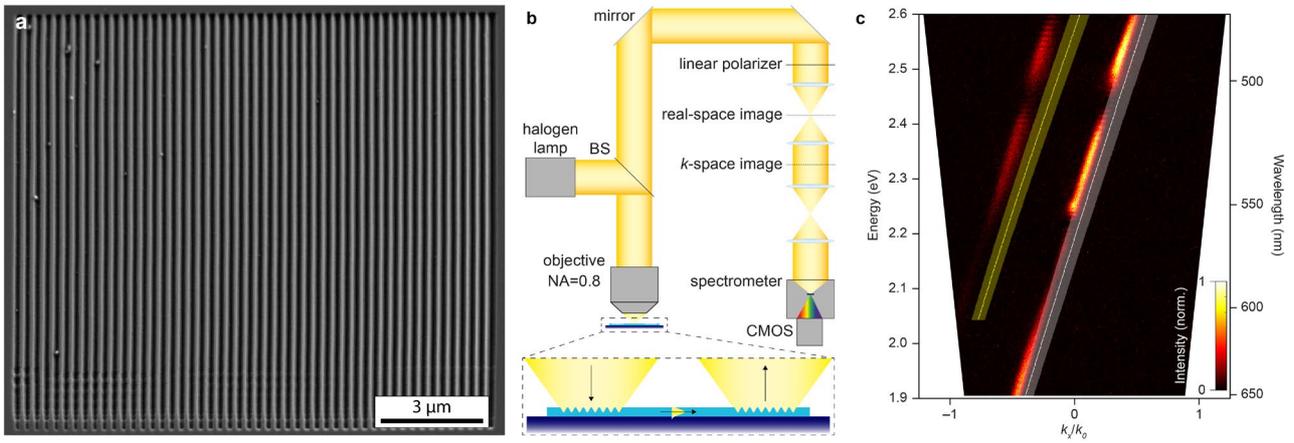

**Extended Data Figure 6 | hBN grating couplers. a**, SEM image of a sinusoidal grating coupler patterned in an hBN flake (255 nm thick). **b**, Schematic of the optical setup used for the guided-mode coupling measurements (details in Methods). The setup is used to couple broadband light into a guided-mode in the hBN with the first grating. The mode then propagates toward the second grating and is outcoupled. The outcoupled light is analyzed in $k$ space (Methods). **c**, Angle-resolved spectra (energy versus in-plane wavevector, $k_x$ with $k_y \approx 0$) for light outcoupled from the second grating, measured using the setup in **b**. The wavevector is relative to $k_0 = 2\pi/\lambda_0$, where $\lambda_0$ is the free-space wavelength of the photon. The red streaks are the experimental data. The white (yellow) line shows the calculated dispersion curve for the $TE_0$ ($TE_1$) mode, diffracted into the light cone by the grating (see Methods). The shaded regions represent a ±3% uncertainty on the reported indices of refraction for hBN (ref. [38]). For all structural design parameters, see Extended Data Table 1.

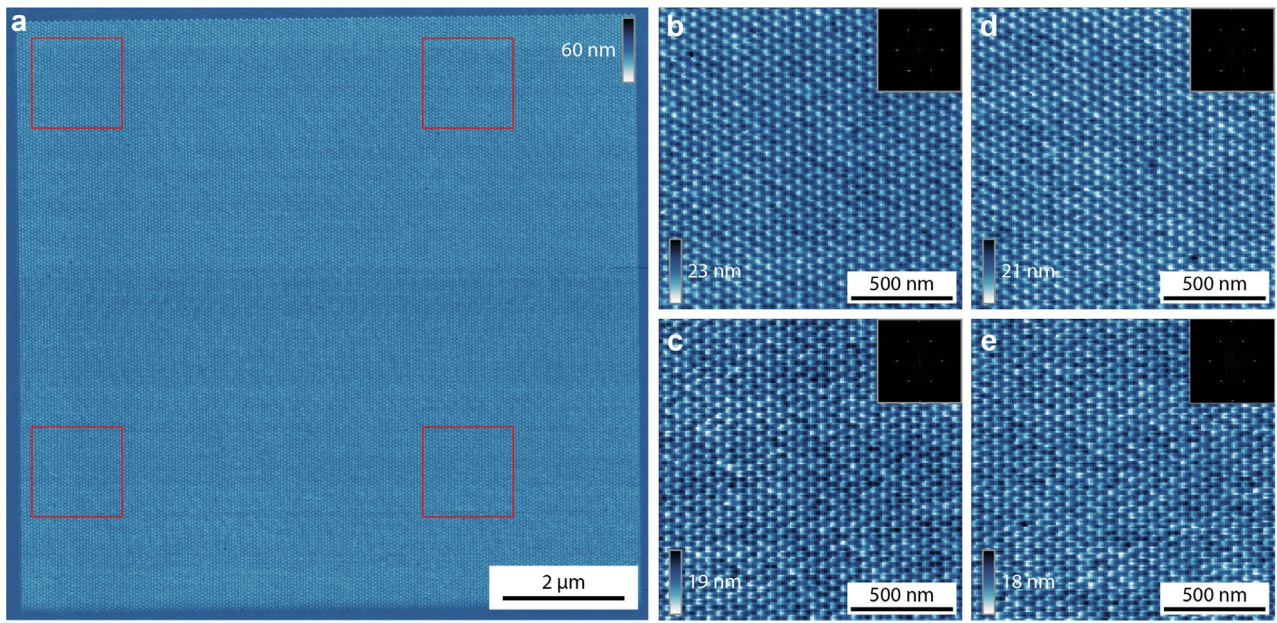

**Extended Data Figure 7 | Large-area electronic Fourier surfaces in hBN. a**, Measured topography (AFM) of a 10×10 µm² hexagonal electronic Fourier surface in hBN, defined with three sinusoids with a period of 50 nm (Methods). **b-e**, Magnified regions from **a**, revealing a high-quality structure over the entire pattern area. The regions are marked with red boxes in **a**. Insets show FFTs for each pattern. For all structural design parameters, see Extended Data Table 1.

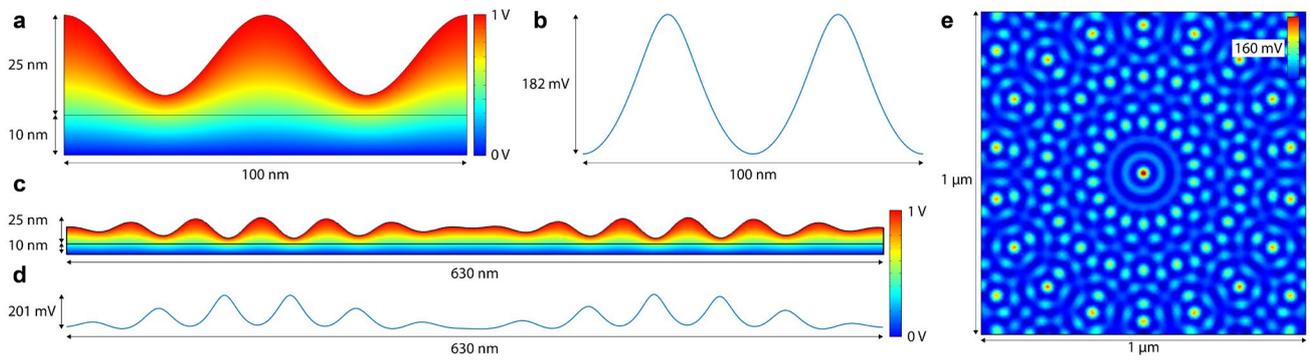

**Extended Data Figure 8 | Electrostatic simulations of a 2D layer surrounded by hBN. a**, Simulation of a single-sinusoidal electronic Fourier surface in hBN with a period of 50 nm. The top hBN has a thickness of 25 nm with a 20 nm depth modulation. The bottom hBN has a thickness of 10 nm and is unstructured. The thin horizontal black line in between the two hBN layers represents the active layer, which could be a monolayer such as $MoS_2$. A voltage of 1 V is applied to the top surface, representing a top gate, and a voltage of 0 V is applied to the bottom surface, representing a back gate. The color map represents the electric potential at every point throughout the structure. **b**, Simulated electric-potential profile at the active layer (thin black horizontal line in **a**) when 1 V is applied to the structure in **a**. The surface profile in the top hBN is revealed in the field profile. **c,d**, As in **a,b**, but for a surface with two sinusoids with periods of 55 and 47 nm. **e**, Simulation of the electric-potential profile at the active layer, as in **b,d**, but for a 3D simulation of the quasicrystal structure from Fig. 4l, defined with 9 sinusoids, each with 50 nm period (Methods). For all structural design parameters, see Extended Data Table 1.